\documentclass{aa}
\usepackage[draft]{hyperref}
\usepackage[varg]{txfonts}
\usepackage{graphicx}
\usepackage[draft]{hyperref}
\usepackage{natbib}
\usepackage{color}
\bibpunct{(}{)}{;}{a}{}{,} % to follow the A&A style
\definecolor{darkgreen}{rgb}{0.0,0.5,0.0}
\definecolor{darkred}{rgb}{0.5,0.0,0.0}
\definecolor{brown}{rgb}{0.65,.16,0.16}
\definecolor{grey}{rgb}{0.4,0.5,0.6}

\begin{document}

% \title{A More Robust Deprojection of the S\'ersic Profile}
\title{A Precise Analytical Approximation for the Deprojection of the S\'ersic Profile}
\titlerunning{Deprojection of the S\'ersic Profile}
\author{Eduardo Vitral\inst{1}
\and Gary A. Mamon\inst{1}
}
\offprints{Eduardo Vitral, \email{vitral@iap.fr}}
\institute{Institut d’Astrophysique de Paris (UMR 7095: CNRS \& Sorbonne Université), 98 bis Bd Arago, F-75014 Paris, France}

\authorrunning{Vitral \& Mamon}
\date{Received; accepted}

\abstract{
% context heading (optional)
      %{
      The S\'ersic model is known to fit well the surface brightness (or surface density) profiles of elliptical galaxies and galaxy bulges, and possibly for dwarf spheroidal galaxies and globular clusters. The deprojected density and mass profiles are important for many astrophysical applications, in particular for mass-orbit modeling of these systems. 
      %}
% aims heading (mandatory)
      %{
      However, the exact deprojection formula for the S\'ersic model employs special functions not available in most computer languages. 
     %}
% results (mandatory)
      %{
We show that all previous analytical approximations to the 3D density profile are imprecise at low S\'ersic index ($n \lesssim 1.5$).
We have derived a more precise analytical approximation to the deprojected S\'ersic density profile by fitting two-dimensional 10th-order polynomials to the differences of the logarithms of the numerical deprojection and of the analytical approximation by \citeauthor{LimaNeto+99} (\citeyear{LimaNeto+99},  LGM) of the density profile on one hand and of the mass profile on the other. Our LGM-based polynomial fits have typical relative precision better than 0.2\% for both density and mass profiles, for S\'ersic indices $0.5 \leq n \leq 10$ and radii $0.001 < r/R_{\rm e} < 1000$. 
      %}
      % conclusions (optional)
        %{
Our 
% LGM-based polynomial
approximation is much more precise than those of 
% PS,  
LGM, \cite{Simonneau&Prada99,Simonneau&Prada04}, \cite{Trujillo+02} for non-half-integer values of the index, and of \cite{Emsellem&vandeVen08}  for non-one-tenth-integer values with $n \lesssim 3$, and are nevertheless more than 0.2\% precise for larger S\'ersic indices, for both density and mass profiles. An appendix compares the deprojected S\'ersic profiles with those of the popular simple models from \cite{Plummer1911}, \cite{Jaffe83}, \cite{Hernquist90}, \cite{Navarro+96}, and \cite{Einasto65}.
}

\keywords{Galaxies: structure -- Galaxies: bulges -- Galaxies: elliptical and
  lenticular, cD -- (Galaxy:) globular clusters: general -- Methods: numerical}
\maketitle

\section{Introduction} 

The Sérsic model \citep{Sersic63,Sersic68} is the generalization of the $R^{1/4}$ law   \citep{deVaucouleurs48} to describe the surface brightness profiles of elliptical galaxies \citep*{Caon+93} and the bulges of spiral galaxies \citep{Simard+11}. It has also been used to describe   the surface density profiles of nuclear star clusters \citep{Carson+15}, resolved dwarf spheroidal galaxies \citep{Battaglia+06} and globular clusters \citep{Barmby+07}.

% \ev{
% \begin{itemize}
%     \item Baumgardt et al.(2009) - The velocity dispersion and mass-to-light ratio of the remote halo globular cluster NGC2419
%     \item McLaughlin\&al08 : \url{https://ui.adsabs.harvard.edu/abs/2008MNRAS.384..563M/abstract}
%     \item Barmby\&al07 : \url{https://ui.adsabs.harvard.edu/abs/2007AJ....133.2764B/abstract}
% \end{itemize}
% }).

The surface (mass or number) density (or equivalently surface brightness) of the S\'ersic model is
\begin{equation}  \label{eq: Sersic}
    \Sigma(R) = \Sigma_{0} \, \exp{\left[ - b_{n} \left(\frac{R}{R_{\mathrm{e}}}\right)^{1/n} \right]} \ ,
\end{equation}
where $R$ is the projected distance to the source center in the \textit{plane-of-sky} (POS), $R_{\mathrm{e}}$ is the effective radius containing half of the projected luminosity, $n$ is the S\'ersic index and $\Sigma_{0}$ is the central surface density. The term $b_{n}$ is a function of $n$, obtained by solving the equation:

\begin{equation} \label{eq: solveb}
    \Gamma(2n)/2 = \gamma(2n,b_{n}) \ ,
\end{equation}
where $\gamma(a,x) = \int_{0}^{x} t^{a-1} e^{-t} \mathrm{d} t$ is the lower incomplete gamma function. 

Since the S\'ersic model represents well astronomical objects viewed in projection, it is important to know its corresponding three-dimensional (3D) density and mass profiles. These serve as a reference to compare to other possible observational tracers, as well as to dark matter. Moreover, the 3D density profile is required for modeling the kinematics of spherical structures, since it appears in the Jeans equation of local dynamical equilibrium. Since the Jeans equation also contains the total mass profile,
the 3D mass profiles of stellar components are required to estimate the dark matter mass profile of elliptical and dwarf spheroidal galaxies. 
% Although a very useful and robust profile, the

% discussion of simple models
Many authors assume that simple three-dimensional models resemble S\'ersic models for certain values of the S\'ersic index: 
It is often assumed that massive ellipticals and spiral bulges are well represented by the \cite{Hernquist90} model (e.g., \citealp{Widrow&Dubinski05}). On the other hand, dwarf spheroidal galaxies are often described as a \cite{Plummer1911} model (e.g. \citealp{Munoz+18}, who also tried S\'ersic and other models), while ultra diffuse galaxies have been described by Einasto  \citep{Einasto65,Navarro+04} models \citep{Nusser19}.
\cite{Lokas&Mamon01} noted that the projected \citeauthor*{Navarro+96} (\citeyear{Navarro+96}, hereafter NFW) model resembles an $n=3$ S\'ersic for reasonable concentrations.
Finally, $n=4$ S\'ersic models are considered to resemble the \cite{Jaffe83} model \citep*{Ciotti+19}.
In Appendix~\ref{sec:comparison}, we compare these models to the deprojected S\'ersic.  

Unfortunately, the deprojection of the S\'ersic surface density profile to a 3D (mass or number)\footnote{The number profile always has the same form as the  mass profile, and is obtained by simply replacing $M(r)$ by $N(r)$ and $M_\infty$ by $N_\infty$, e.g. in Eqs.~(\ref{eq: MPSofr}), (\ref{eq: MPS}), and (\ref{eq: Mofr}).}
density profile, through \cite{Abel1826} inversion 
\begin{equation} \label{eq: Abelinv}
    \rho(r) = - \frac{1}{\pi} \int_{r}^{+\infty} \frac{\mathrm{d} \Sigma}{\mathrm{d} R} \frac{\mathrm{d} R}{\sqrt{R^{2}-r^{2}}} \ ,
\end{equation}
(e.g., \citealp{Binney&Mamon82}),
as well as the corresponding 3D mass (or number) profile 
\begin{equation}
    M(r) = \int_{0}^{r} 4 \pi s^{2} \rho(s) \, \mathrm{d} s \ ,
\end{equation}
both involve the complicated Meijer~G special function (\citealp{Mazure&Capelato02} for integer values of $n$, and \citealp{Baes&Gentile11} for general values of $n$) or the other complicated Fox $H$ function \citep{Baes&vanHese11},
% \begin{equation}
% \nu(r)     
% \end{equation}
neither of which are available in popular computer languages.

% \ev{Baes \& Gentile 2010 and Baes \& van Hese 2011 seem to have perfected this expression, which was only available for $n$ integers, and now is extended to all values of $n$. Baes \& Ciotti 2019 also targeted small values of $n$, but did not focus on the same things we did.}

% considerably complicated deprojection into volume density, which can be calculated, by considering spherical symmetry and applying an Abel inversion, along with an also complicated mass profile:

% \citet{Mazure&Capelato02} have calculated an analytical solution for those integrals, which depends on Meijer G functions, which are non-trivial functions that can increase computation time and slow fitting procedures for example, making it a very demanding solution. That is why

Following the shape of the analytical approximation to the $R^{1/4}$ law by \cite{Mellier&Mathez87},
\citeauthor{Prugniel&Simien97} (\citeyear{Prugniel&Simien97}, hereafter, PS) proposed  an analytical approximation for the 3D density of the S\'ersic profile:

\begin{equation} \label{eq: rhoPS}
    \rho_{\mathrm{PS}}(r) = \rho_{0} \left(\frac{r}{R_{\mathrm{e}}}\right)^{-p_{n}} \exp{\left[- b_{n} \left(\frac{r}{R_{\mathrm{e}}}\right)^{1/n}\right]} \ ,
\end{equation}
which yields a simple analytical form for the 3D mass profile 
% \begin{equation} \label{eq: mPS}
%     M_{\mathrm{PS}}(r) = 4 \pi \, \rho_{0} \, R_{\mathrm{e}}^{3} \, n \, b_{n}^{n(p_{n}-3)} \, \gamma \left(n(3-p_{n}),b_{n} \left(\frac{r}{R_{\mathrm{e}}}\right)^{1/n}\right) \ ,
% \end{equation}
\begin{eqnarray}
M_{\rm PS}(r) &\!\!\!\!=\!\!\!\!& M_\infty \,{\gamma[(3-p_n)\,n,b_n\,(r/R_{\rm e})^{1/n}]\over \Gamma[(3-p_n)\,n]} \ ,
\label{eq: MPSofr}\\
M_\infty &\!\!\!\!=\!\!\!\!& 4\pi\,\rho_0\,R_{\rm e}^3\,{n\,\Gamma[(3-p_n)\,n]\over b_n^{(3-p_n)\,n}} \ ,
\label{eq: MPS}
\end{eqnarray}
 where $p_{n}$ is a function depending only on $n$. PS calculated this dependence to be:
\begin{equation}  \label{eq: pPS}
    p_{n,\mathrm{PS}} = 1 - {0.594\over n} + {0.055 \over n^2} \ ,
\end{equation}
while 
 \citet*{LimaNeto+99} (hereafter,  LGM) later perfected this approximation with
\begin{equation}  \label{eq: rhoLGM}
    p_{n,\mathrm{LGM}} = 1 - {0.6097\over n} + {0.05463 \over n^2} \ .
\end{equation}
LGM indicate that equation~(\ref{eq: rhoLGM}) is good to 5 per cent relative accuracy for $0.56 \leq n \leq 10$ and $-2 < \log(r/R_{\rm e}) < 3$.
However, the power-law approximation at small radii is unjustified for small $n$. Indeed, as shown by \cite{Baes&Gentile11}, the central density profile converges to a finite value for $n<1$ (and the inner density profile diverges only logarithmically for $n=1$), as we will illustrate in Sect.~\ref{sec:results}.

\citeauthor{Simonneau&Prada99} (\citeyear{Simonneau&Prada99}, \citeyear{Simonneau&Prada04}, hereafter SP) proposed
the quasi-Gaussian expansion for the density profile
\begin{equation}
   \rho_{\rm SP}(r) = {2\over \pi} {b_n\over (n\!-\!1)}{\Sigma_0\over R_{\rm e}}\left({r\over R_{\rm e}}\right)^{1/n-1}\!\sum_{j=1}^5\rho_j\,\exp\left[-b_n\lambda_j\left({r\over R_{\rm e}}\right)^{1/n} 
\right] ,
\label{eq: rhoSP}
\end{equation}
where 
\begin{eqnarray}
    \lambda_j&=& \left(1-x_j^2\right)^{-1/(n-1)} \ ,\\
    \rho_j &=& w_j\,{x_j\over \sqrt{1-\left(1-x_j^2\right)^{2n/(n-1)}}} \ ,
\end{eqnarray}
where $x_j$ and $w_j$ are 10 fit parameters.
The individual SP density profiles (the terms inside the sum of Eq.~\ref{eq: rhoSP}) have a similar (but not the same) form as the PS/LGM one, hence a similar shape for the mass profile:
%\begin{multline}
\begin{eqnarray}
   M_{\rm SP}(r) &\!\!\!\!=\!\!\!\!& M_{\infty} \, {4 \over \pi \, (n-1) \,
     \Gamma(2n)} \nonumber \\ &\!\!\!\!\mbox{}\!\!\!\!& \times \sum_{j=1}^5 {\rho_j \over \lambda_j^{2n+1}} \, \gamma\left[2n+1,b_n\lambda_j\left({r\over R_{\rm e}}\right)^{1/n} 
\right] \ .
\label{eq: MSP}    
\end{eqnarray}

%\end{multline}
 
\cite{Trujillo+02} proposed an ellipsoid formula, which in the limit of spherical symmetry becomes 
\begin{equation} \label{eq: Trapprox}
\rho_{\rm T}(r) = {2^{(n-1)/(2n)}\, b_n\over \pi\, n}
\,{\Sigma_0\over R_{\rm e}}
\,r^{p_n(1/n-1)} 
 {
 K_{\nu_n}\left(r/ R_{\rm e}\right)
 \over 1-\sum_{i=0}^2 a_{n,i} \log^i \left(r/R_{\rm e}\right)
 } \ ,
\end{equation}
where $K_\nu(x)$ is the modified Bessel function of the 2nd kind\footnote{\cite{Trujillo+02} call this the modified Bessel function of the 3rd kind, as some others do.}  of index $\nu$, while $\nu_n$, $p_n$, $a_{n,0}$, $a_{n,1}$, and $a_{n,2}$ are empirical functions of index $n$. 
\citeauthor{Trujillo+02} only provided their results for integer and half-integer values of $n$ for $n \leq 5$ and only integer values of $n$ beyond.
\begin{figure}
\centering
\includegraphics[width=\hsize]{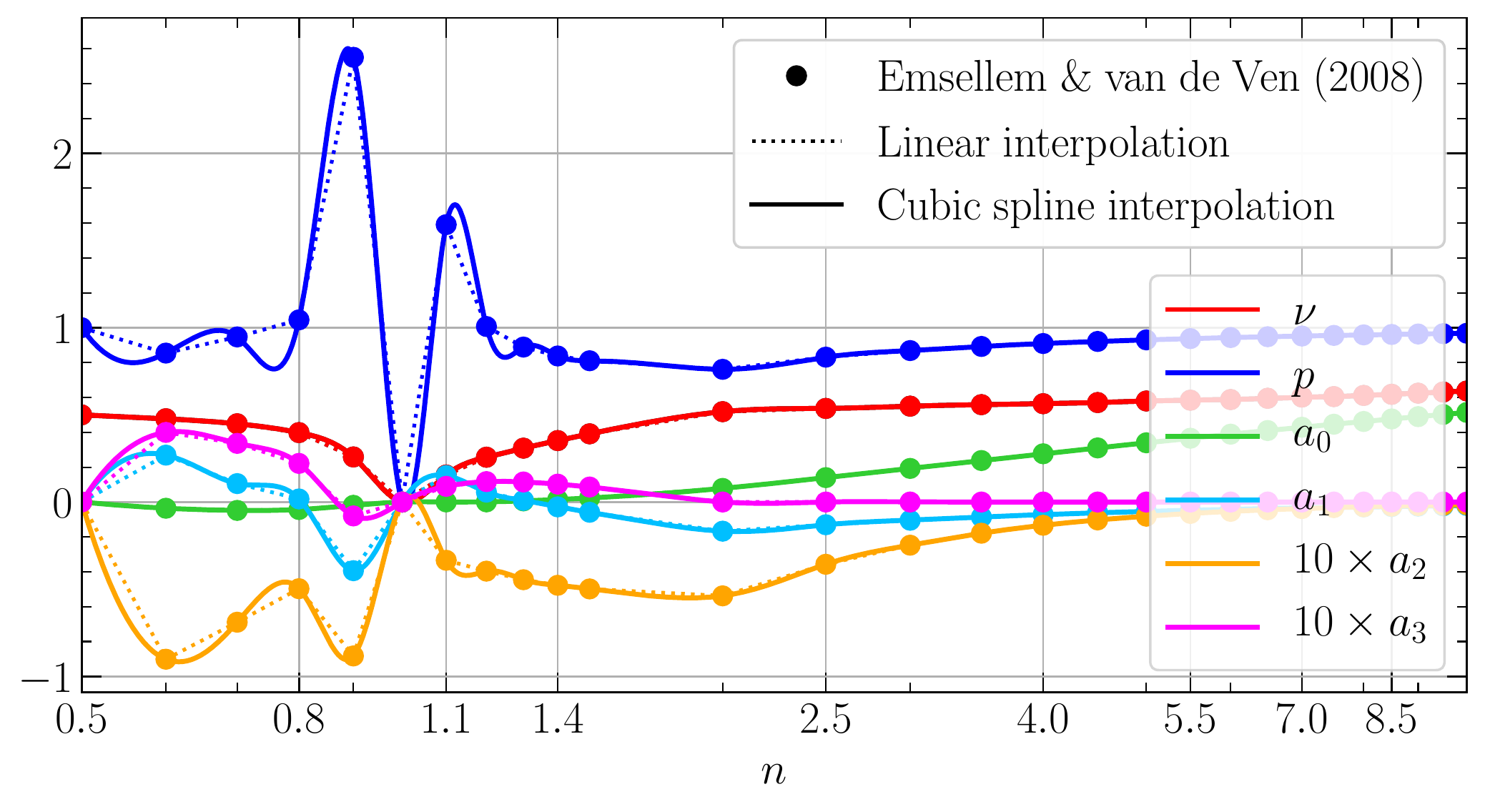}
%\raggedright
%\centering
\caption{Variation with S\'ersic index of the different parameters of the analytical approximation of \cite{Emsellem&vandeVen08} for the deprojected S\'ersic density profile (\emph{filled circles}). The \emph{solid} and \emph{dotted} curves show the spline cubic and linear interpolations, respectively. At small $n$, the parameters vary abruptly and the interpolations (both linear and cubic) are thus uncertain.
\label{fig: spline}}
\end{figure}
\citeauthor{Emsellem&vandeVen08} (\citeyear{Emsellem&vandeVen08}, hereafter EV) repeated their analysis
 on a finer grid of $n$, with steps of 0.1 for $0.5 \leq n \leq 1.5$ and with one more term, $a_{n,3}$, in equation~ (\ref{eq: Trapprox}), involving 168 parameters. Unfortunately, as shown in Figure~\ref{fig: spline}, these functions vary abruptly for $n\lesssim1.2$. Moreover, neither \citeauthor{Trujillo+02} nor \citeauthor{Emsellem&vandeVen08} provide analytical forms for the mass profile. 
 %\ev{I would even say that they vary abruptly for $n<2$. Also, we should cite Emsellem \& van de Ven 08.}
 
In summary, all the previous approximations to the deprojected S\'ersic model have drawbacks: 
\begin{itemize}
%\vspace{-\baselineskip}
    \item PS, LGM (and \citealp{Marquez+00}, which is the same as LGM, but with a slightly different last term for $p_n$, which was a typo) are inappropriate for low $n$ \citep{Baes&Gentile11} and less precise than claimed \citep{Emsellem&vandeVen08}.
    \item SP is limited to $n\geq 1$, and is generally less precise than EV.
    \item \cite{Trujillo+02} is only given for half-integer values of $n$ and their parameters vary wildly with $n$ for $n\leq 1.5$. They do not provide a formula for the mass profile.
    \item EV also suffers from discrete values of $n$, even though the grid is finer ($\Delta n = 0.1$ for $n\leq 1.5$). EV also did not provide a formula for the mass profile.
\end{itemize}

In this article, we 
 provide polynomial fits to the log residuals of the LGM approximation, allowing to reach high accuracy for both the 3D density and 3D mass profiles in a wide range of S\'ersic indices.
In Sect.~\ref{sec:method}, we present the mathematical formalism and briefly explain our numerical integration method.
We then show  in Sect.~\ref{sec:results} how our polynomial plus LGM approximation is orders of magnitude more precise than the formulae of LGM, SP, and \cite{Trujillo+02}, as well as that of  EV for low $n$ and only slightly worse for $n \ga 3$.  We conclude and discuss our results in Sect.~\ref{sec:conclude}.

\section{Method}
\label{sec:method}

\subsection{Equations using dimensionless profiles}

We express the general surface density, 3D density, and 3D mass (or number) profiles in terms of dimensionless functions:
\begin{eqnarray} \label{eq: tilderho}
\Sigma(R) &\!\!\!\!=\!\!\!\!& {M_\infty\over \pi R_{\rm e}^2}\,
\widetilde \Sigma \left ({R\over R_{\rm e}}\right) \ , \\
\rho(r) &\!\!\!\!=\!\!\!\!& \left(\frac{M_{\infty}}{4 \pi R_{\mathrm{e}}^3}\right) \,
        \widetilde\rho
        \left({r\over R_{\rm e}}\right) \ , \\
 M(r) &\!\!\!\!=\!\!\!\!& M_{\infty} \,
        \widetilde M
        \left({r\over R_{\rm e}}\right) \ .
        \label{eq: Mofr}
\end{eqnarray}
%\ev{I don't agree with the $-1$ in the exponent in the formula above.}
%
Hereafter, we will use $x = r/R_{\rm e}$ and $X = R/R_{\rm e}$. For the S\'ersic model, the dimensionless surface density profile is 
\begin{equation}
    \widetilde \Sigma_{\rm S}(X) = {b_n^{2n} \over 2n\,\Gamma(2n)}\,\exp \left (-b_n\, X^{1/n}\right) \ ,
\end{equation}
while for the PS model, one can write the dimensionless 3D density and mass profiles as 
\begin{eqnarray}
    \widetilde \rho_{\rm PS}(x) &\!\!\!\!=\!\!\!\!&  \frac{b_{n}^{(3-p_n)\,n}}{n \, \Gamma\left[(3-p_n)\,n\right]} \, x^{-p_{n}} \, \exp{\left[-b_{n} x^{1/n}\right]} \ ,\\
    \widetilde M_{\rm PS}(x) &\!\!\!\!=\!\!\!\!& {\gamma\left[(3-p_n)\,n, b_n \,X^{1/n}\right] \over \Gamma((3-p_n)\,n)}\ .
\end{eqnarray}

It is easy to show that the deprojection equation~(\ref{eq: Abelinv}) becomes
\begin{equation} \label{eq: tildeAbelinv}
    \widetilde{\rho}(x) = - \frac{4}{\pi} \int_{x}^{+\infty} \frac{\mathrm{d} \widetilde{\Sigma}}{\mathrm{d} X} \frac{\mathrm{d} X}{\sqrt{X^{2}-x^{2}}} \ ,
\end{equation}
where 
\begin{equation}
    {{\rm d}\widetilde\Sigma \over {\rm d}X} \equiv \widetilde\Sigma'(X)
    = -{b_n^{2n+1}\over 2n^2\,\Gamma(2n)}\,X^{-1+1/n}\,
    \exp{\left(-b_{n} x^{1/n}\right)} \ .
    \label{eq: rhotildeofx}
\end{equation}
The dimensionless mass profile is
\begin{eqnarray}
    \widetilde M(x) &\!\!\!\!=\!\!\!\!&
    \int_0^x y^2\, \widetilde\rho(y)\,{\rm d}y 
    = -{4\over \pi}\,\int_0^x y^2\,{\rm d}y\,\int_y^\infty {\widetilde\Sigma'(X)\over \sqrt{X^2-y^2}}\,{\rm d}X
    \label{eq: Mofx1}
    \\
    &\!\!\!\!=\!\!\!\!&
   - \int_0^x X^2\,\widetilde \Sigma'(X)\,{\rm d}X
    -{2\over \pi}\,
    \int_x^\infty X^2\,\sin^{-1}\left ({x\over X}\right)\,\widetilde\Sigma'(X)\,{\rm d}X \nonumber \\
    &\!\!\!\!\mbox{}\!\!\!\!& +{2\over \pi}\,
    \int_x^\infty x\,\sqrt{X^2-x^2}\,\widetilde\Sigma'(X)\,{\rm d}X\ ,
    \label{eq: Mtildeofxbis}
\end{eqnarray}
where  equation~(\ref{eq: Mtildeofxbis}) is obtained by inversion of the order of integration in the second equality of  Eq.~(\ref{eq: Mofx1}).

\subsection{Numerical integration}
We numerically evaluated the dimensionless 3D density (Eq. [\ref{eq: tildeAbelinv}]) and mass (Eq.~[\ref{eq: Mtildeofxbis}]) profiles by 
% integrating equation~(\ref{eq: Abelinv})
peforming the numerical integrations 
in cells 50 $\times$ 100 of $[\log n, \log(r/R_{\rm e})]$, with $\log 0.5 \leq \log n \leq 1$ and $-3 \leq \log(r/R_{\rm e}) \leq 3$.
Numerical calculations were done with Python's \textsc{scipy.integrate.quad}.
For both density and mass profiles, we split the numerical integration in two, i.e.
\begin{equation}
    \int_a^b f(X) \,{\rm d}X = \int_a^{X_{\rm crit}} f(X)\,{\rm d}X
    + \int_{X_{\rm crit}}^b f(X)\,{\rm d}X  \ ,
\end{equation}
where $\exp\left(-b_n X_{\rm crit}^{1/n}\right) = 10^{-9}$ and $a \leq X_{\rm crit} \leq b$. 
We used a relative tolerance of ${\tt epsrel}=10^{-4}$ and ${\tt limit}=1000$ in both integrals. If $X_{\rm crit} \not \in [a,b]$, we also used ${\tt epsrel}=10^{-4}$ and  ${\tt limit}=1000$, but for a single integral from $a$ to $b$.

We performed our analysis using either the highly accurate approximations for $b_n$ of \citeauthor{Ciotti&Bertin99} (\citeyear{Ciotti&Bertin99}, hereafter, CB) or the exact (numerical) solutions of Eq.~(\ref{eq: solveb}). We noticed that the difference between these two approaches was negligible (see Sect.~\ref{sec:results}).

We then fit two-dimensional polynomials to  both $\log{\left[\widetilde{\rho}_{\mathrm{LGM}}(x,n)/\widetilde{\rho}(x,n)\right]}$ and $\log{\left[\widetilde{M}_{\mathrm{LGM}}(x,n)/\widetilde{M}(x,n)\right]}$, 
for geometrically spaced $x$ and $n$,  i.e. writing
\begin{equation}    \label{eq: poly}
    \log{\left[\widetilde{f}_{\mathrm{LGM}}/\widetilde{f}\right]} = - \sum_{i=0}^{k}\sum_{j = 0}^{k-i} a_{ij} \log^{i}x \,\log^{j}n
\end{equation}
with polynomial orders $2 \leq k \leq 12$. 
For this, we used Python's package \textsc{numpy.linalg.lstsq}. 
% \ev{I think we should mention that the smaller residuals were found for $n=10$.}
We found the smallest residuals for order~10 polynomials when using both the $b_n$ approximation of CB and $b_n$ by numerically solving Eq.~(\ref{eq: solveb}). The coefficients are provided in Tables~\ref{tab: coeffnu}, \ref{tab: coeffM}, \ref{tab: coeffnub} and \ref{tab: coeffMb} in Appendix~\ref{sec:coeffs}. In the rest of the paper, we present the results relative to the CB approximation, since it is a simpler and more used model, and also because our order~10 polynomial fits remarkably well the exact $b_n$ case.

\subsection{Numerical precision: Tests for known simple analytical deprojections ($n=0.5$ and $1$)}

For S\'ersic indices $n=0.5$ and $n=1$, 
there are analytical solutions for the 3D density profile:
%\begin{eqnarray}   
\begin{equation}
    \widetilde{\rho}(x) = \left\{
    \begin{array}{ll}
    4\,
    \displaystyle \frac{b_{0.5}^{3/2}}{ \sqrt{\pi}} \, \exp{\left[-b_{0.5} \, x^2\right]}
    & \qquad (n=0.5) \ ,
    \label{eq: n05}\\
    \\
%\label{eq: n1}
\displaystyle
    2\, \frac{ b_1^3}{\pi} \, K_0(b_1 x) & \qquad (n=1) \ ,
\end{array} 
\right.
\end{equation}
where $K_0(x)$ is the modified Bessel function of the second kind of index 0.
We can therefore
  verify the numerical integration of  Eq.~(\ref{eq: tildeAbelinv}) for these two S\'ersic indices.

For the interval $-3 \leq \log(r/R_{\rm e}) \leq 3$, we compared the densities from numerical integration with the analytical formulae of  Eq.~(\ref{eq: n05}), using the CB approximation for $b_n$. The match is very good, with \emph{root-mean-square} (rms) values of  $\log\left(\widetilde{\rho}_{\rm ana}/\widetilde{\rho}_{\rm num}\right)$ of $1.5 \times 10^{-7}$ and $2 \times 10^{-8}$ for $n=0.5$ and $n=1$, respectively. The same comparison using the exact $b_n$ yields $7 \times 10^{-5}$ and $2 \times 10^{-8}$, respectively (with one particular value of $r$ causing the higher rms for $n=0.5$).

\section{Results}
\label{sec:results}

\begin{figure}
\centering
\includegraphics[width=\hsize]{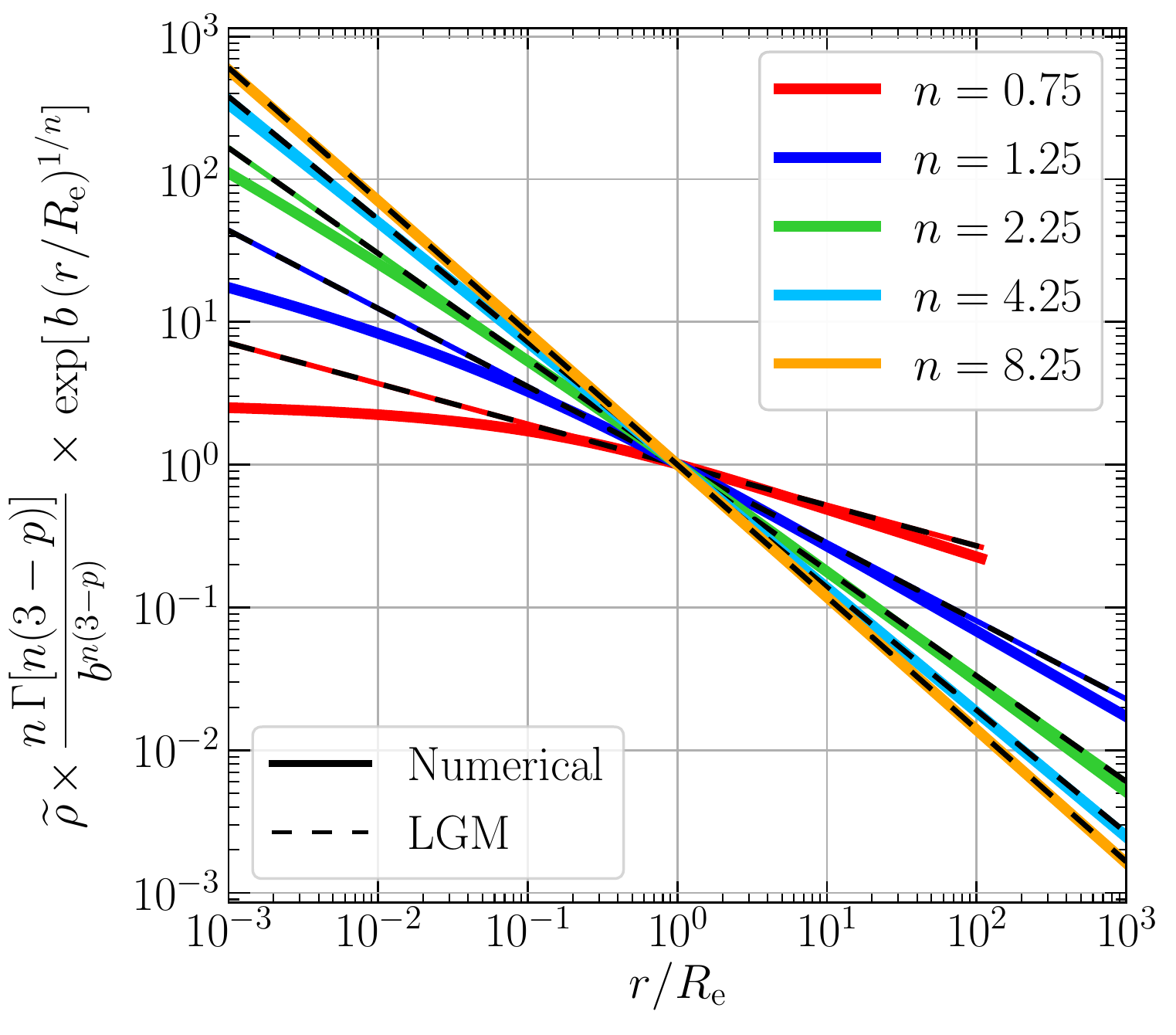}
%\raggedright
%\centering
\caption{Illustration of the accuracy of the PS formula with the LGM coefficients for $p_n$. The \emph{solid curves} show the numerically estimated profiles, while the \emph{colored-dashed curves} show the LGM approximation.
\label{fig: rhoexpXx}}
\end{figure}

As seen in Figure~\ref{fig: rhoexpXx}, the 3D density profiles depart from the power laws proposed by LGM at low $n$, especially for low radii, as expected by the asymptotic expansions of \cite{Baes&Gentile11} for $n<1$. Interestingly, the LGM formula is also inadequate at low radii for $n=1.25$ and 2.25, although the asymptotic expansion of \citeauthor{Baes&Gentile11} indicate power-law behavior at small radii.
This poor accuracy of the LGM approximation at low radii is a serious concern when performing kinematic modeling of systems with possible central massive black holes. For example, Gaia DR2 positions and proper motions for stars in nearby globular clusters extend inwards to 0.7 arcsec from the center, which translates to $0.002\,R_{\rm e}$.

\begin{figure*}
\centering
\includegraphics[width=\hsize]{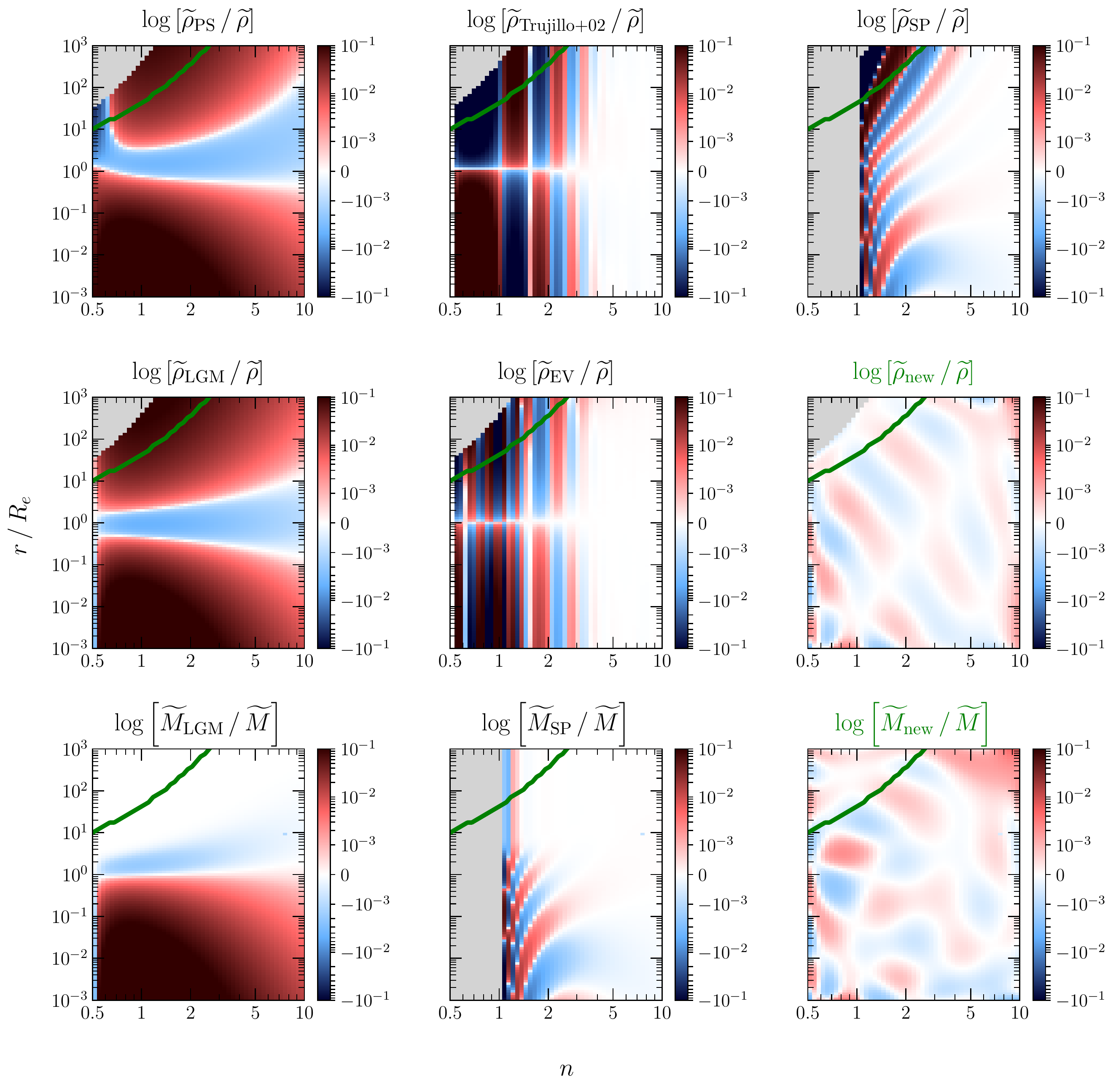} 
\vspace{0.2cm}
\caption{
Accuracy of deprojected density (\emph{top 6 panels}) and mass (\emph{3 bottom panels}) of the different analytical approximations (LGM: Lima-Neto et al. 1999; SP: Simonneau \& Prada 1999, 2004; Trujillo+02: Trujillo et al. 2002; EV: Emsellem \& van de Ven 2008;   our new one  (Eq.~[\ref{eq: fnew}], with green-colored titles) as a function of both S\'ersic index (abscissae) and radii (ordinates).
The color scale given in the \emph{vertical color bars} are linear for log ratios between --0.001 and 0.001 and logarithmic beyond. The \emph{gray region} and \emph{green curves}  in the upper left of the density panels are for regions where the numerical integration reached the underflow limit or density $10^{-30}$ times $\rho(R_{\rm e})$, respectively, because of the very rapid decline of density at large radii for low $n$, and also covers $n<1$ that is not covered by the SP model. Note that the EV and Trujillo+02 models perform better at specific values of $n$ that are often missed in our grid.
\label{fig: grid}}
\end{figure*}

%We now compare the accuracy of the different analytical approximations for the 3D density and 3D mass profiles, omitting PS and \citeauthor{Trujillo+02}, since LGM and EV were designed to improve on the respective previous two models.

We now compare the accuracy of the different analytical approximations for the 3D density and 3D mass profiles. Figure~\ref{fig: grid} displays the ratio $\log \left( \widetilde{f}_{\mathrm{model}} \, / \, \widetilde{f} \right)$, for $\widetilde f = \widetilde \rho$ and $\widetilde f = \widetilde M$, for the main analytical approximations available in the literature, along with our new model 
\begin{equation}
%f_{\rm new}\left ({r\over R_{\rm e}},n\right) = 
%f_{\rm LGM}\left ({r\over R_{\rm e}},n\right) \,
%{\rm dex}\left[\sum_{i=0}^{10}\sum_{j=0}^{10-i} a_{i,j}\log^i \left({r\over %R_{\rm e}}\right)\,\log^j n \right] ,
f_{\rm new}(x,n) = f_{\rm LGM}(x,n) \ {\rm dex}\left[\sum_{i=0}^{10}\sum_{j=0}^{10-i} a_{i,j}\log^i x\,\log^j n \right] \ ,
\label{eq: fnew}
\end{equation}
where $f$ is either the  3D density or 3D mass profile.
We see that our model presents a more continuous behavior over the full range of S\'ersic indices and radii. Our approximation displays the smallest residuals among all models for $n\la3$ (except that SP outperforms our model for mass estimates at $r>3 \,R_{\rm e}$ for $n>1.3$).
% (i.e. $n\lesssim3$). 

\begin{figure*}
\centering
\includegraphics[width=0.75\hsize]{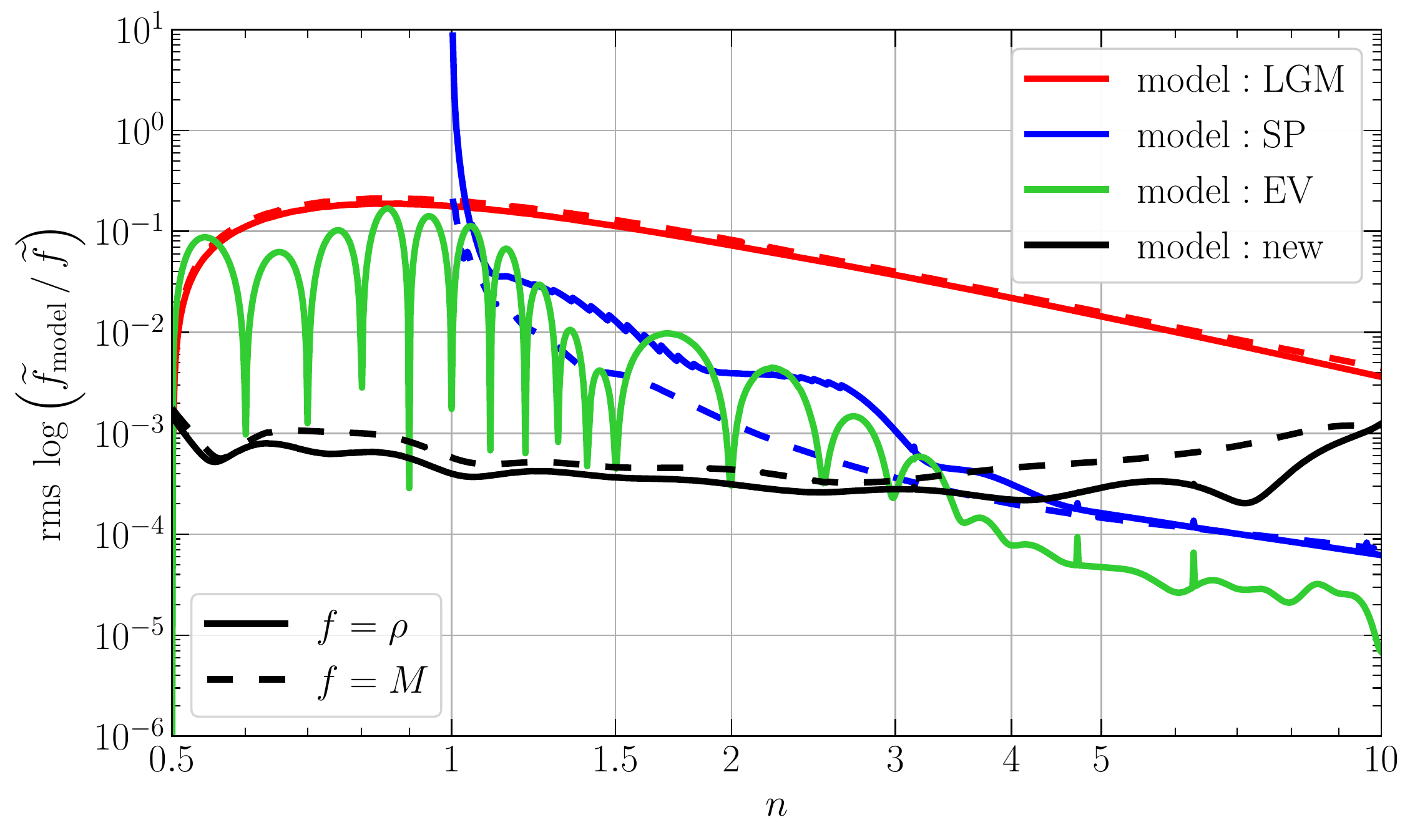}
%\raggedright
%\centering
\caption{
Accuracy of different approximations (LGM: Lima-Neto et al. 1999; SP: Simonneau \& Prada 1999, 2004; EV: Emsellem \& van de Ven 2008;  our new one  (Eq.~[\ref{eq: fnew}]) as a function of S\'ersic index.
Note that the EV model performs better at specific values of $n$ that are often missed in our logarithmic grid of 1000 values of $n$. 
\label{fig: rms}}
\end{figure*}

The variation of accuracy with S\'ersic index can  be seen in more detail in Figure~\ref{fig: rms}, which displays the rms of $\log \left(\widetilde{f}_{\mathrm{model}} \, / \, \widetilde{f}\right)$, over the radial domain where $\rho(r) > 10^{-30}\,\rho(R_{\rm e})$, of the main analytical approximations, using 1000 log-spaced S\'ersic indices.
Figure~\ref{fig: rms} indicates that the SP (respectively, EV) approximation for density has rms relative accuracy worse than 2.3\% (0.01 for $\log \left(\widetilde{f}_{\mathrm{model}} \, / \, \widetilde{f}\right)$) for $n < 1.6$ (respectively 1.3).
% Figure~\ref{fig: rms} shows that our 
Our
approximation (Eq.~[\ref{eq: fnew}]) is more accurate than SP for $n < 4.3$ (density)
and
$n < 3.1$ (mass),
and is more accurate than EV for
$n < 3.4$ (density), except for their particular choices of $n$.
Figure~\ref{fig: rms} shows that the EV approximation is much more accurate at specific values of $n$ (note that our grid does not contain all of these values precisely, so the EV approximation is even more accurate at these specific values of $n$).  
However, these specific values of $n$ represent a negligible measure compared to the full continuous range of $0.5 \leq n \leq 10$. Therefore, the EV approximation at low $n$ is not reliable for estimating the 3D density profile.

% It is important to mention that 
We analyzed the results shown in Figure~\ref{fig: rms} using $b_n$ from either CB or by numerically solving  Eq.~(\ref{eq: solveb}), and the results were very similar. 
In fact, the results are similar if we adopt one form of $b_n$ in the numerical integration and the other in the analytical approximations.
% Moreover, we confirm that even if one wants to use our order~10 polynomial calculated for $b_{n,\, \rm CB}$ in an analysis that adopts $b_n$ by solving  Eq.~(\ref{eq: solveb}), the results are extremely similar. 
This can be explained by the fact that  $\log \left(\widetilde{f}_{\mathrm{LGM}} \, / \, \widetilde{f}\right)$ is practically the same for both estimates of $b_n$, yielding a very similar fit of  Eq.~(\ref{eq: poly}).

Finally, we provide in Table~\ref{tab: rms} the rms accuracies computed over the full range of radii $-3 \leq \log(r/R_{\rm e}) \leq 3$ and $0.5 \leq n \leq 10$, except for the SP formula, which does not allow $n \leq 1$, and also avoiding the domain where $\rho(r) < 10^{-30}\,\rho(R_{\rm e})$. We see that, averaging over all S\'ersic indices, our approximation is much more accurate than all others (with over 10 times lower rms).

\begin{table}
\centering
\caption{Accuracy of approximations to 3D density and mass profiles}
\label{tab: rms}
\begin{center}
\tabcolsep=2.5pt
\begin{tabular}{lcc} 
  \hline\hline
  & rms & rms \\
Author & $\log\left(\displaystyle {\widetilde{\rho}_{\mathrm{approx}}\over \widetilde{\rho}_{\mathrm{num}}}\right)$
 & 
$\log\left(\displaystyle{\widetilde{M}_{\mathrm{approx}}\over \widetilde{M}_{\mathrm{num}}}\right)$
\\
\hline                      
    Prugniel \& Simien 97
     & 0.1052 & 0.1187  \\     
 %   \cite{LimaNeto+99
    Lima Neto et al. 99  & 0.0905 & 0.1021  \\
    %\cite{Simonneau&Prada04} 
    Simonneau \& Prada 04 ($n> 1$)& 0.0238 & 0.0098  \\
    % \cite{Trujillo+02} 
    Trujillo et al. 02 & 0.1496 & --  \\
    % \cite{Emsellem&vandeVen08}
    Emsellem \& van de Ven 08 & 0.0382 &-- \\
    {\bf new} & {\bf 0.0005} & {\bf 0.0007} \\
    {\bf hybrid-1} (optimized for $n_{\rm cut}$) & {\bf 0.0004} & {\bf 0.0005}  \\
    {\bf hybrid-2} (optimized for $r_{\rm cut}$) & {\bf 0.0004} & {\bf 0.0005}  \\
\hline    
\end{tabular}
\end{center}

\parbox{\hsize}{Notes: The rms accuracies are computed over the full range of radii $-3 \leq \log(r/R_{\rm e}) \leq 3$ (100 steps) and $0.5 \leq n \leq 10$ (50 steps), except for the SP formula, which does not allow $n \leq 1$, and also avoiding the domain where $\rho(r) < 10^{-30}\,\rho(R_{\rm e})$. \cite{Trujillo+02} and EV do not provide analytical mass profiles. 
The lower two rows display hybrid models, both with our new approximation $\widetilde{\rho}_{\rm new}$ for $n<3.4$ and $\widetilde{\rho}_{\rm EV}$ for $n\geq3.4$. The first hybrid model has a mass profile $\widetilde{M}_{\rm new}$ for $n<3$ and $\widetilde{M}_{\rm SP}$ for $n\geq3$, while in hybrid model 2, the mass profile is $\widetilde{M}_{\rm new}$ for $r<R_{\rm e}$ and $\widetilde{M}_{\rm LGM}$ for $r\geq R_{\rm e}$. 
}
\end{table}

% \subsection{Simonneau et al. (2004) approximation}

% \subsection{Trujillo et al. (2002) approximation}

%\gam{We may end up merging the subsections!}

\section{Conclusions and Discussion}
\label{sec:conclude}

The S\'ersic model is usually considered to provide excellent fits to the surface density (or surface brightness) profiles of elliptical galaxies, spiral bulges, and even dwarf spheroidal galaxies and globular clusters. In the past, many authors have used simple analytical models to describe these systems, arguing that their models, once projected,   resemble S\'ersic models. It is more relevant to compare the physically meaningful three-dimensional density profiles of these simple models to the deprojected S\'ersic model. 

This comparison is made in Appendix~\ref{sec:comparison} for the Plummer, Jaffe, Hernquist, Einasto, and NFW models. As seen in Fig.~\ref{fig: other-models}, most of the simple models do not provide decent fits to the deprojected S\'ersic model, even for narrow ranges of the S\'ersic index. The Plummer model requires a low index at small radii, but a much higher index at large radii, and the normalized density profile fits poorly at most radii. The Hernquist model resembles the $n=2.8$ deprojected S\'ersic model at low radii and the $n=5.7$ S\'ersic at large radii.
The NFW models resemble the $n=2.8$ deprojected S\'ersic at low radii (consistent with the similarity of the projected S\'ersic with NFW discovered by \citealp{Lokas&Mamon01}), but have a shallower slope at large radii than even the shallowest ($n=8$) deprojected S\'ersic model. 
On the other hand, the Jaffe model resembles the $n=5.7$ model at all radii. 
Moreover, as seen in Figure~\ref{fig: Einasto}, the Einasto model provides a fair representation (rms difference of density profiles normalized to value at half-mass radius less than 0.1 dex) of the deprojected S\'ersic model for
%$0.5 < n < 0.51$ and
$n > 6.5$.

We reconsidered the different analytical deprojections of the S\'ersic surface brightness (or surface density) profile.
We found that the analytical approximations present in the literature do not show satisfying results when the S\'ersic index is in the range $0.5 \leq n \lesssim 1.5$ (apart from the specific values of $n$ given by \citealp{Emsellem&vandeVen08}). 
In particular, the power-law times exponential density profile of \cite{Prugniel&Simien97} and \cite{LimaNeto+99} fails to reproduce the inner density profiles for low $n$, even up to $n=2.25$ despite the power-law behavior expected at small radii for $n>1$ \citep{Baes&Gentile11}.  

With an order~10 two-dimensional polynomial fit, we propose a new analytical approximation (Eq.~[\ref{eq: fnew}])
that is precise over 
the range $\log 0.5 \leq \log n \leq 1$, for $-3 \leq \log \left(r/R_{\mathrm{e}}\right) \leq 3$.
Our approximation 
% of Eq.~(\ref{eq: fnew}) 
provides the highest precision when averaging over all values of S\'ersic indices and radii (Table~\ref{tab: rms}).
While the approximations of \cite{Simonneau&Prada99,Simonneau&Prada04} on one hand and of \cite{Emsellem&vandeVen08} on the other, are more accurate than ours for $n > 4.3$ and 3.4, respectively, ours is more accurate at lower S\'ersic indices.

This is important for the study of astronomical sources with low S\'ersic indexes, such as galaxy bulges, nuclear star clusters, dwarf spheroidal galaxies, and globular clusters.
Moreover, our approximation of Eq.~(\ref{eq: fnew}) to the density profile is sufficiently accurate for most scientific analyses for $n>3$.
But the user could use a hybrid approximation, combining either the \citeauthor{Simonneau&Prada04} or \citeauthor{Emsellem&vandeVen08}
 approximations for $n\geq3.4$ and ours for $n < 3.4$ (as shown in the last rows of Table~\ref{tab: rms}). 
Finally, our analysis has the advantage of also providing a precise approximation for the mass profile, whereas no analytical expression can be derived from the density profile of  \citeauthor{Emsellem&vandeVen08}.
Our Python~3 codes are available at \url{https://github.com/eduardo-vitral/Vitral_Mamon2020a} along with coefficients of Tables~\ref{tab: coeffnu} and \ref{tab: coeffM}.

These results will be useful in future mass-orbit modeling analyses of low-mass spherical systems, as we are preparing for globular clusters (Vitral \& Mamon, in prep.).

\bibliography{master}

\begin{appendix}
%\onecolumn

\section{Comparison of deprojected S\'ersic to other popular models}
\label{sec:comparison}

\begin{figure}
\centering
\includegraphics[width=\hsize]{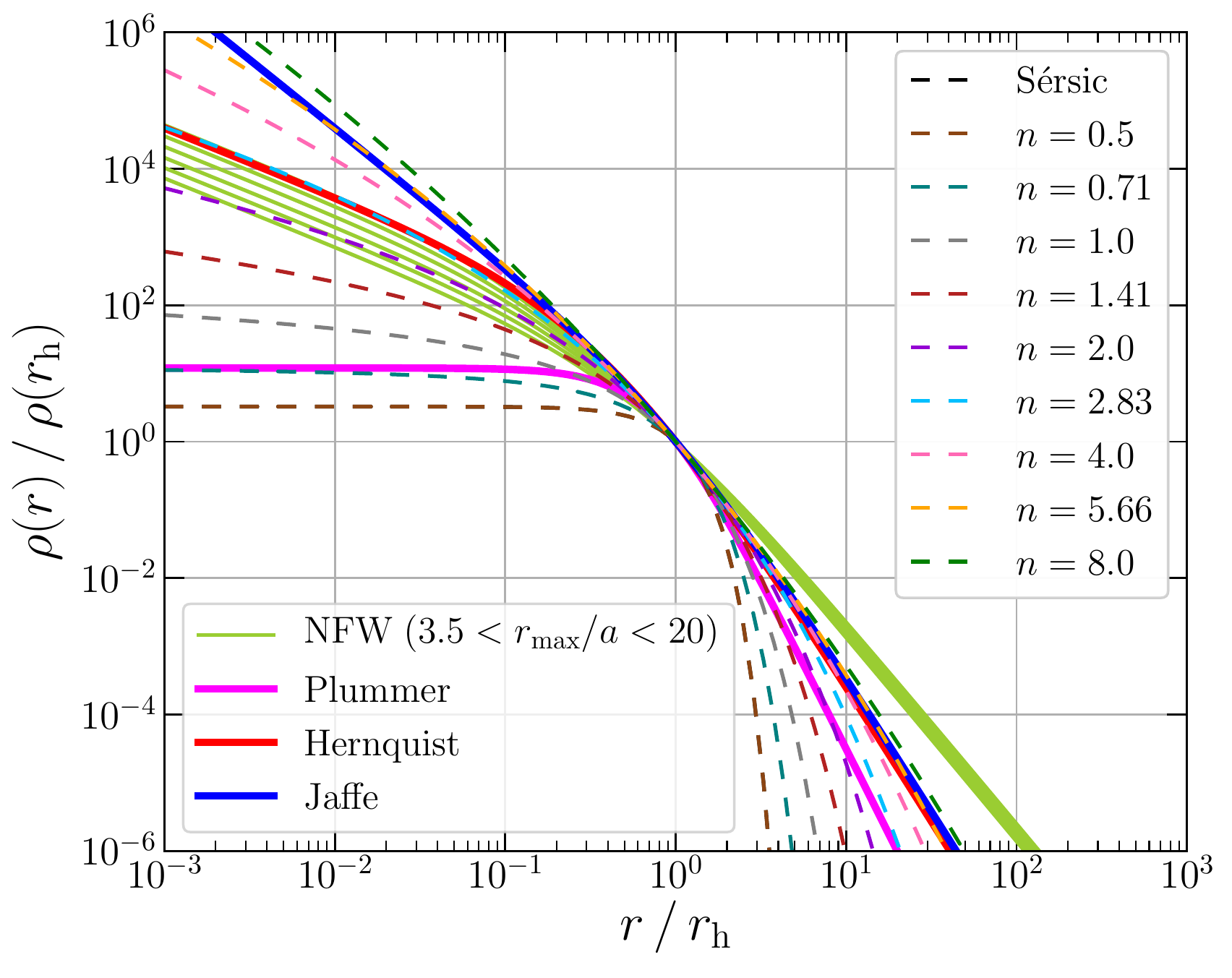}
%\raggedright
%\centering
\caption{Comparison of other known three-dimensional density profiles and the deprojected S\'ersic density profile for certain values of the S\'ersic index $n$. All density profiles are normalized to the value at the three-dimensional half-mass radius, $r_{\rm h}$ (see text). The different NFW models can be distinguished at low radii, where the density increases with $r_{\rm max}/a$.
\label{fig: other-models}}
\end{figure}

% Here, we provide the expressions used for popular three-dimensional density models in order to plot Figure~\ref{fig: other-models}, where we compare them with the S\'ersic profile for certain values of the S\'ersic index $n$.
Figure~\ref{fig: other-models} compares the density profiles, normalized to the half-mass radius $r_{\rm h}$, for which we applied the following relations: 
\begin{equation}
    \rho(r) \propto \left\{
    \def\arraystretch{2.5}
    \begin{array}{ll}
    \displaystyle \left[1 + \left({r \over a}\right)^2\right]^{-5/2} & \rm (Plummer) \ , \\
    \displaystyle  \left({r \over a}\right)^{-2} \left[1 + \left({r \over a}\right)\right]^{-2} & \rm (Jaffe) \ ,\\
    \displaystyle \left({r \over a}\right)^{-1} \left[1 + \left({r \over a}\right)\right]^{-3}
    & \rm (Hernquist) \ ,\\
    \displaystyle \left({r \over a}\right)^{-1} \left[1 + \left({r \over a}\right)\right]^{-2}
    & \rm (NFW) \ ,\\
    \displaystyle \exp\left[-\left ({r\over a}\right)^{1/n}\right] & \rm (Einasto) \ .
    \end{array}
    \right.
\end{equation}
The ratio of half-mass radius to scale radius $r_{\rm h}/a$ is given by
\begin{equation}
    {r_{\rm h}\over a} = \left\{
      \def\arraystretch{1.5}
    \begin{array}{ll}
 \left[\left(1+2^{1/3}\right)/\sqrt{3}\right]\        &  \rm (Plummer)\ ,\\
1 & \rm (Jaffe) \ ,\\
 \left(1 + \sqrt{2}\right)        & \rm (Hernquist) \ ,\\
{\rm dex}\left(-0.209 + 0.856\,\log c-0.090 \log^2 c \right)\!\!\!& \rm (NFW), \\
 \left[P^{(-1)}(3n,1/2)\right]^n & \rm (Einasto) \ .
     \end{array}
\right.
\label{eq: rhalf}
\end{equation}
In Eq.~(\ref{eq: rhalf}) for NFW,  $c = r_{\rm max}/a$, where $r_{\rm max}$ is the maximum allowed radius (because, contrary to all other models discussed here, the NFW model has logarithmically divergent mass).
Also, for Einasto, $P^{(-1)}(a,y)$ is the inverse regularized lower incomplete gamma function, i.e. $x = P^{(-1)}(a,y)$ satisfies $\gamma(a,x)/\Gamma(a)=y$.\footnote{The inverse (regularized) incomplete gamma function is available in many computer languages, e.g. Python (\textsc{scipy} package), Fortran, Matlab, Mathematica, and Javascript.}
     For S\'ersic, the conversion was done by fitting a order 3 polynomial and recovering the relation $r_{\rm h}/R_{\rm e} = \!\sum_{j=0}^3 a_i \, \log^i n$, where $\{a_0, \, a_1, \, a_2, \, a_3 \} = \{1.32491, \, 0.0545396, \, -0.0286632, \, 0.0035086 \}$.

% \begin{itemize}
% \itemsep 0.5\baselineskip
%     \item Plummer: $r_{\rm h} = \left[\left(1+2^{1/3}\right)/\sqrt{3}\right]\,a$ 
%     \item Hernquist: $r_{\rm h} = \left(1 + \sqrt{2}\right) \, a$.
%     \item Jaffe: $r_{\rm h} = a$.
%     \item NFW: $r_{\rm h} = c \, a$, where 

    % Einasto: $r_{\rm h} = \left[P^{(-1)}(3n,1/2)\right]^n\,a$, where 
%\end{itemize}

% \begin{equation}
%     \mathrm{Plummer \ profile: } \
%     \rho(r) \propto \left[1 + \left({r \over a}\right)^2\right]^{-5/2}  \ .
% \end{equation}

% \begin{equation}
%     \mathrm{Jaffe \ profile: } \
%     \rho(r) \propto \left({r \over a}\right)^{-2} \left[1 + \left({r \over a}\right)\right]^{-2} \ .
% \end{equation}

% \begin{equation}
%     \mathrm{Hernquist \ profile: } \
%     \rho(r) \propto \left({r \over a}\right)^{-1} \left[1 + \left({r \over a}\right)\right]^{-3} \ .
% \end{equation}

% \begin{equation}
%     \mathrm{NFW \ profile: } \
%     \rho(r) \propto \left({r \over a}\right)^{-1} \left[1 + \left({r \over a}\right)\right]^{-2} \ .
% \end{equation}

% \begin{equation}
%     \mathrm{Einasto \ profile: } \
%     \rho(r) \propto \exp\left[-\left ({r\over a}\right)^{1/n}\right] \ .
% \end{equation}

\begin{figure}[ht]
    \centering
    \includegraphics[width=\hsize]{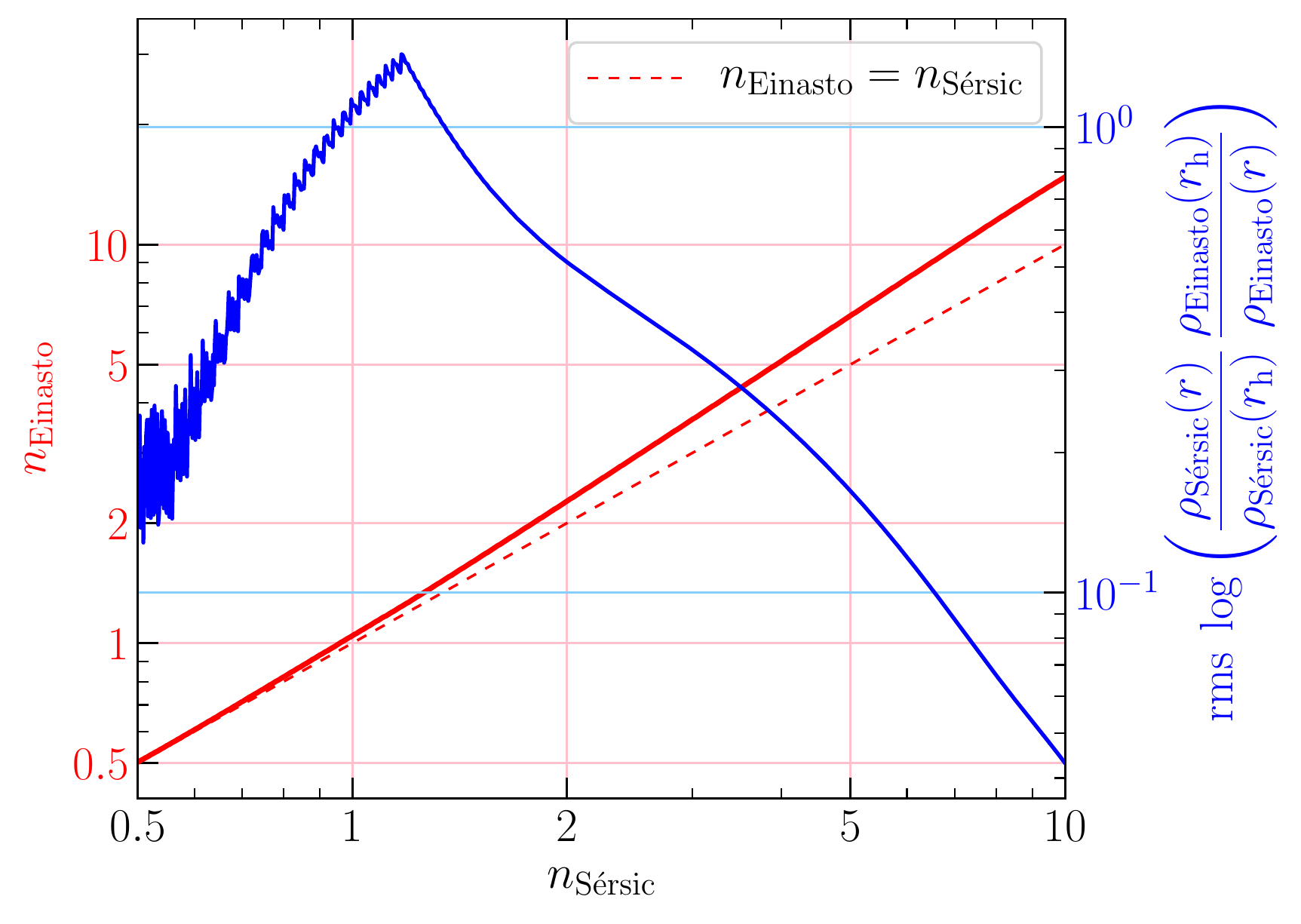}
    \caption{Comparison of Einasto and deprojected S\'ersic density profiles (both normalized to half-mass radius). {\bf Red}: best-fit Einasto index (dashed line is $n_{\rm Einasto}=n_{\rm S\'ersic}$). {\bf Blue}: rms of best fit.
    \label{fig: Einasto}}
\end{figure}

The Einasto model, which is the 3D analog of the S\'ersic model, resembles the deprojected S\'ersic model. 
Figure~\ref{fig: Einasto} shows the best-fit values of the Einasto index, $n_{\rm Einasto}$ in terms of the S\'ersic index. The relation (red curve) is almost one-to-one (dashed line). The figure also shows the rms over all radii and best-fit indices (blue curve).

\section{Coefficients of polynomials for new deprojected S\'ersic models}
\label{sec:coeffs}
In this section, we present Tables~\ref{tab: coeffnu}, \ref{tab: coeffM}, \ref{tab: coeffnub} and \ref{tab: coeffMb}, containing the coefficients $a_{ij}$ in  Eq.~(\ref{eq: poly}), for $\widetilde{f} = \widetilde{\rho}$ and $\widetilde{f} = \widetilde{M}$, as well as for both \cite{Ciotti&Bertin99} approximation for $b_n$ and the exact solution from  Eq.~(\ref{eq: solveb}). The numbers in parentheses are the exponents: e.g. ``$1.234 \, (-3)$'' corresponds to $1.234 \times 10^{-3}$. Coefficients that are not followed by a number in parentheses have exponent zero. 
%\ev{Do you know any command to force all the tables in the same page?}

\begin{table*}[ht]
\centering
\caption{Coefficients of  equations (\ref{eq: poly}) and (\ref{eq: fnew}), for $\widetilde{f} = \widetilde{\rho}$ and $b_n$ calculated from \cite{Ciotti&Bertin99} accurate approximation.} 

\label{tab: coeffnu}
%\centering                         
\tabcolsep=2.4pt
\footnotesize
\begin{tabular}{rrrrrrrrrrrr}
\hline \hline
$i$\textbackslash$j$& \multicolumn{1}{c}{0} & \multicolumn{1}{c}{1} & \multicolumn{1}{c}{2} & \multicolumn{1}{c}{3} & \multicolumn{1}{c}{4} & \multicolumn{1}{c}{5} & \multicolumn{1}{c}{6} & \multicolumn{1}{c}{7} & \multicolumn{1}{c}{8} & \multicolumn{1}{c}{9} & \multicolumn{1}{c}{10} \\
\hline
0 & $ 5.017 \, (-3)$ & $ 1.573 \, (-3)$ & $-7.175 \, (-2)$ & $ 1.256 \, (-1)$ & $ 4.047 \, (-1)$ & $-1.191 \ \ \ \ \ \ \ \, $ & $ 2.455 \, (-1)$ & $ 8.650 \, (-1)$ & $ 1.178 \ \ \ \ \ \ \ \, $ & $-2.771$ & $ 1.209$ \\
1 & $-4.507 \, (-3)$ & $-8.623 \, (-3)$ & $ 5.430 \, (-2)$ & $ 2.298 \, (-1)$ & $-9.349 \, (-1)$ & $-3.113 \, (-1)$ & $ 5.052 \ \ \ \ \ \ \ \, $ & $-7.980 \ \ \ \ \ \ \ \, $ & $ 5.065 \ \ \ \ \ \ \ \, $ & $-1.162$ & \multicolumn{1}{c}{--} \\
2 & $-4.251 \, (-2)$ & $ 6.921 \, (-2)$ & $ 2.706 \, (-1)$ & $-1.134 \ \ \ \ \ \ \ \, $ & $ 1.318 \ \ \ \ \ \ \ \, $ & $ 7.078 \, (-1)$ & $-3.093 \ \ \ \ \ \ \ \, $ & $ 2.664 \ \ \ \ \ \ \ \, $ & $-7.604 \, (-1)$ & \multicolumn{1}{c}{--} & \multicolumn{1}{c}{--} \\
3 & $ 1.373 \, (-2)$ & $-2.444 \, (-2)$ & $-8.324 \, (-2)$ & $ 2.795 \, (-1)$ & $-1.923 \, (-1)$ & $-2.219 \, (-1)$ & $ 3.670 \, (-1)$ & $-1.381 \, (-1)$ & \multicolumn{1}{c}{--} & \multicolumn{1}{c}{--} & \multicolumn{1}{c}{--} \\
4 & $ 1.428 \, (-3)$ & $-7.563 \, (-3)$ & $ 2.897 \, (-4)$ & $ 5.385 \, (-2)$ & $-1.248 \, (-1)$ & $ 1.200 \, (-1)$ & $-4.310 \, (-2)$ & \multicolumn{1}{c}{--} & \multicolumn{1}{c}{--} & \multicolumn{1}{c}{--} & \multicolumn{1}{c}{--} \\
5 & $-1.388 \, (-3)$ & $ 3.261 \, (-3)$ & $ 1.615 \, (-3)$ & $-1.112 \, (-2)$ & $ 1.272 \, (-2)$ & $-5.059 \, (-3)$ & \multicolumn{1}{c}{--} & \multicolumn{1}{c}{--} & \multicolumn{1}{c}{--} & \multicolumn{1}{c}{--} & \multicolumn{1}{c}{--} \\
6 & $-9.613 \, (-5)$ & $ 5.901 \, (-4)$ & $-4.636 \, (-4)$ & $-4.867 \, (-4)$ & $ 4.043 \, (-4)$ & \multicolumn{1}{c}{--} & \multicolumn{1}{c}{--} & \multicolumn{1}{c}{--} & \multicolumn{1}{c}{--} & \multicolumn{1}{c}{--} & \multicolumn{1}{c}{--} \\
7 & $ 9.505 \, (-5)$ & $-1.801 \, (-4)$ & $ 1.074 \, (-5)$ & $ 5.734 \, (-5)$ & \multicolumn{1}{c}{--} & \multicolumn{1}{c}{--} & \multicolumn{1}{c}{--} & \multicolumn{1}{c}{--} & \multicolumn{1}{c}{--} & \multicolumn{1}{c}{--} & \multicolumn{1}{c}{--} \\
8 & $ 4.306 \, (-6)$ & $-2.496 \, (-5)$ & $ 2.464 \, (-5)$ & \multicolumn{1}{c}{--} & \multicolumn{1}{c}{--} & \multicolumn{1}{c}{--} & \multicolumn{1}{c}{--} & \multicolumn{1}{c}{--} & \multicolumn{1}{c}{--} & \multicolumn{1}{c}{--} & \multicolumn{1}{c}{--} \\
9 & $-2.924 \, (-6)$ & $ 4.229 \, (-6)$ & \multicolumn{1}{c}{--} & \multicolumn{1}{c}{--} & \multicolumn{1}{c}{--} & \multicolumn{1}{c}{--} & \multicolumn{1}{c}{--} & \multicolumn{1}{c}{--} & \multicolumn{1}{c}{--} & \multicolumn{1}{c}{--} & \multicolumn{1}{c}{--} \\
10 & $-2.117 \, (-8)$ & \multicolumn{1}{c}{--} & \multicolumn{1}{c}{--} & \multicolumn{1}{c}{--} & \multicolumn{1}{c}{--} & \multicolumn{1}{c}{--} & \multicolumn{1}{c}{--} & \multicolumn{1}{c}{--} & \multicolumn{1}{c}{--} & \multicolumn{1}{c}{--} & \multicolumn{1}{c}{--} \\
\hline   
\end{tabular}
\normalsize
%\parbox{\hsize}{\textit{Notes}: The coefficient $a_{ij}$ is placed on the row $i$ and column $j$ 
% (rows and columns go from 0 to 10). 
%The numbers in parentheses are the exponents: e.g. ``$1.234 \, (-3)$'' corresponds to $1.234 \times 10^{-3}$. Coefficients that are not followed by a number in parentheses have exponent zero.}
\end{table*}
\vspace{-3mm}
\begin{table*}[ht]
\caption{Coefficients of  equation (\ref{eq: poly}) and (\ref{eq: fnew}), for $\widetilde{f} = \widetilde{M}$ and $b_n$ calculated from \cite{Ciotti&Bertin99} accurate approximation. }    
\label{tab: coeffM}
\centering                      
\tabcolsep=2.4pt
\footnotesize
\begin{tabular}{rrrrrrrrrrrr}
\hline \hline
$i$\textbackslash$j$& \multicolumn{1}{c}{0} & \multicolumn{1}{c}{1} & \multicolumn{1}{c}{2} & \multicolumn{1}{c}{3} & \multicolumn{1}{c}{4} & \multicolumn{1}{c}{5} & \multicolumn{1}{c}{6} & \multicolumn{1}{c}{7} & \multicolumn{1}{c}{8} & \multicolumn{1}{c}{9} & \multicolumn{1}{c}{10} \\
\hline
0 & $ 7.076 \, (-4)$ & $ 1.014 \, (-4)$ & $-3.794 \, (-2)$ & $ 2.785 \, (-2)$ & $ 4.247 \, (-1)$ & $-6.904 \, (-1)$ & $-6.342 \, (-1)$ & $ 8.367 \, (-1)$ & $ 2.565 \ \ \ \ \ \ \ \, $ & $-4.115$ & $ 1.622$ \\
1 & $ 1.557 \, (-2)$ & $-4.107 \, (-2)$ & $-6.114 \, (-2)$ & $ 7.489 \, (-1)$ & $-1.700 \ \ \ \ \ \ \ \, $ & $-3.157 \, (-2)$ & $ 5.508 \ \ \ \ \ \ \ \, $ & $-8.413 \ \ \ \ \ \ \ \, $ & $ 5.055 \ \ \ \ \ \ \ \, $ & $-1.080$ & \multicolumn{1}{c}{--} \\
2 & $-3.517 \, (-2)$ & $ 5.740 \, (-2)$ & $ 2.328 \, (-1)$ & $-9.638 \, (-1)$ & $ 1.007 \ \ \ \ \ \ \ \, $ & $ 1.001 \ \ \ \ \ \ \ \, $ & $-3.228 \ \ \ \ \ \ \ \, $ & $ 2.706 \ \ \ \ \ \ \ \, $ & $-7.776 \, (-1)$ & \multicolumn{1}{c}{--} & \multicolumn{1}{c}{--} \\
3 & $ 1.986 \, (-2)$ & $-2.017 \, (-2)$ & $-1.276 \, (-1)$ & $ 3.000 \, (-1)$ & $-8.746 \, (-2)$ & $-4.227 \, (-1)$ & $ 5.202 \, (-1)$ & $-1.817 \, (-1)$ & \multicolumn{1}{c}{--} & \multicolumn{1}{c}{--} & \multicolumn{1}{c}{--} \\
4 & $ 1.274 \, (-3)$ & $-4.467 \, (-3)$ & $ 1.989 \, (-3)$ & $ 3.040 \, (-2)$ & $-7.734 \, (-2)$ & $ 7.555 \, (-2)$ & $-2.723 \, (-2)$ & \multicolumn{1}{c}{--} & \multicolumn{1}{c}{--} & \multicolumn{1}{c}{--} & \multicolumn{1}{c}{--} \\
5 & $-2.853 \, (-3)$ & $ 3.494 \, (-3)$ & $ 8.720 \, (-3)$ & $-1.959 \, (-2)$ & $ 1.656 \, (-2)$ & $-6.196 \, (-3)$ & \multicolumn{1}{c}{--} & \multicolumn{1}{c}{--} & \multicolumn{1}{c}{--} & \multicolumn{1}{c}{--} & \multicolumn{1}{c}{--} \\
6 & $-1.620 \, (-4)$ & $ 1.411 \, (-4)$ & $-4.970 \, (-5)$ & $-3.006 \, (-4)$ & $ 2.779 \, (-4)$ & \multicolumn{1}{c}{--} & \multicolumn{1}{c}{--} & \multicolumn{1}{c}{--} & \multicolumn{1}{c}{--} & \multicolumn{1}{c}{--} & \multicolumn{1}{c}{--} \\
7 & $ 2.557 \, (-4)$ & $-3.086 \, (-4)$ & $-2.993 \, (-4)$ & $ 3.039 \, (-4)$ & \multicolumn{1}{c}{--} & \multicolumn{1}{c}{--} & \multicolumn{1}{c}{--} & \multicolumn{1}{c}{--} & \multicolumn{1}{c}{--} & \multicolumn{1}{c}{--} & \multicolumn{1}{c}{--} \\
8 & $ 1.608 \, (-5)$ & $-2.203 \, (-6)$ & $-3.249 \, (-7)$ & \multicolumn{1}{c}{--} & \multicolumn{1}{c}{--} & \multicolumn{1}{c}{--} & \multicolumn{1}{c}{--} & \multicolumn{1}{c}{--} & \multicolumn{1}{c}{--} & \multicolumn{1}{c}{--} & \multicolumn{1}{c}{--} \\
9 & $-9.523 \, (-6)$ & $ 1.303 \, (-5)$ & \multicolumn{1}{c}{--} & \multicolumn{1}{c}{--} & \multicolumn{1}{c}{--} & \multicolumn{1}{c}{--} & \multicolumn{1}{c}{--} & \multicolumn{1}{c}{--} & \multicolumn{1}{c}{--} & \multicolumn{1}{c}{--} & \multicolumn{1}{c}{--} \\
10 & $-6.227 \, (-7)$ & \multicolumn{1}{c}{--} & \multicolumn{1}{c}{--} & \multicolumn{1}{c}{--} & \multicolumn{1}{c}{--} & \multicolumn{1}{c}{--} & \multicolumn{1}{c}{--} & \multicolumn{1}{c}{--} & \multicolumn{1}{c}{--} & \multicolumn{1}{c}{--} & \multicolumn{1}{c}{--} \\
\hline   
\end{tabular}
\end{table*}
\normalsize
\vspace{-3mm}
\begin{table*}[ht]
\caption{Coefficients of equation (\ref{eq: poly}) and (\ref{eq: fnew}), for $\widetilde{f} = \widetilde{\rho}$ and $b_n$ calculated from  Eq.~(\ref{eq: solveb}).}    
\label{tab: coeffnub}
\centering                      
\tabcolsep=2.4pt
\footnotesize
\begin{tabular}{rrrrrrrrrrrr}
\hline \hline
$i$\textbackslash$j$& \multicolumn{1}{c}{0} & \multicolumn{1}{c}{1} & \multicolumn{1}{c}{2} & \multicolumn{1}{c}{3} & \multicolumn{1}{c}{4} & \multicolumn{1}{c}{5} & \multicolumn{1}{c}{6} & \multicolumn{1}{c}{7} & \multicolumn{1}{c}{8} & \multicolumn{1}{c}{9} & \multicolumn{1}{c}{10} \\
\hline
0 & $ 5.017 \, (-3)$ & $ 1.573 \, (-3)$ & $-7.176 \, (-2)$ & $ 1.256 \, (-1)$ & $ 4.048 \, (-1)$ & $-1.191 \ \ \ \ \ \ \ \, $ & $ 2.460 \, (-1)$ & $ 8.647 \, (-1)$ & $ 1.178 \ \ \ \ \ \ \ \, $ & $-2.771$ & $ 1.209$ \\
1 & $-4.506 \, (-3)$ & $-8.634 \, (-3)$ & $ 5.434 \, (-2)$ & $ 2.298 \, (-1)$ & $-9.353 \, (-1)$ & $-3.103 \, (-1)$ & $ 5.051 \ \ \ \ \ \ \ \, $ & $-7.979 \ \ \ \ \ \ \ \, $ & $ 5.065 \ \ \ \ \ \ \ \, $ & $-1.162$ & \multicolumn{1}{c}{--} \\
2 & $-4.251 \, (-2)$ & $ 6.921 \, (-2)$ & $ 2.706 \, (-1)$ & $-1.134 \ \ \ \ \ \ \ \, $ & $ 1.318 \ \ \ \ \ \ \ \, $ & $ 7.077 \, (-1)$ & $-3.093 \ \ \ \ \ \ \ \, $ & $ 2.664 \ \ \ \ \ \ \ \, $ & $-7.605 \, (-1)$ & \multicolumn{1}{c}{--} & \multicolumn{1}{c}{--} \\
3 & $ 1.373 \, (-2)$ & $-2.443 \, (-2)$ & $-8.325 \, (-2)$ & $ 2.795 \, (-1)$ & $-1.922 \, (-1)$ & $-2.220 \, (-1)$ & $ 3.670 \, (-1)$ & $-1.381 \, (-1)$ & \multicolumn{1}{c}{--} & \multicolumn{1}{c}{--} & \multicolumn{1}{c}{--} \\
4 & $ 1.428 \, (-3)$ & $-7.564 \, (-3)$ & $ 2.912 \, (-4)$ & $ 5.385 \, (-2)$ & $-1.247 \, (-1)$ & $ 1.200 \, (-1)$ & $-4.310 \, (-2)$ & \multicolumn{1}{c}{--} & \multicolumn{1}{c}{--} & \multicolumn{1}{c}{--} & \multicolumn{1}{c}{--} \\
5 & $-1.388 \, (-3)$ & $ 3.260 \, (-3)$ & $ 1.617 \, (-3)$ & $-1.112 \, (-2)$ & $ 1.272 \, (-2)$ & $-5.060 \, (-3)$ & \multicolumn{1}{c}{--} & \multicolumn{1}{c}{--} & \multicolumn{1}{c}{--} & \multicolumn{1}{c}{--} & \multicolumn{1}{c}{--} \\
6 & $-9.613 \, (-5)$ & $ 5.900 \, (-4)$ & $-4.633 \, (-4)$ & $-4.872 \, (-4)$ & $ 4.045 \, (-4)$ & \multicolumn{1}{c}{--} & \multicolumn{1}{c}{--} & \multicolumn{1}{c}{--} & \multicolumn{1}{c}{--} & \multicolumn{1}{c}{--} & \multicolumn{1}{c}{--} \\
7 & $ 9.505 \, (-5)$ & $-1.801 \, (-4)$ & $ 1.072 \, (-5)$ & $ 5.735 \, (-5)$ & \multicolumn{1}{c}{--} & \multicolumn{1}{c}{--} & \multicolumn{1}{c}{--} & \multicolumn{1}{c}{--} & \multicolumn{1}{c}{--} & \multicolumn{1}{c}{--} & \multicolumn{1}{c}{--} \\
8 & $ 4.306 \, (-6)$ & $-2.496 \, (-5)$ & $ 2.464 \, (-5)$ & \multicolumn{1}{c}{--} & \multicolumn{1}{c}{--} & \multicolumn{1}{c}{--} & \multicolumn{1}{c}{--} & \multicolumn{1}{c}{--} & \multicolumn{1}{c}{--} & \multicolumn{1}{c}{--} & \multicolumn{1}{c}{--} \\
9 & $-2.924 \, (-6)$ & $ 4.229 \, (-6)$ & \multicolumn{1}{c}{--} & \multicolumn{1}{c}{--} & \multicolumn{1}{c}{--} & \multicolumn{1}{c}{--} & \multicolumn{1}{c}{--} & \multicolumn{1}{c}{--} & \multicolumn{1}{c}{--} & \multicolumn{1}{c}{--} & \multicolumn{1}{c}{--} \\
10 & $-2.113 \, (-8)$ & \multicolumn{1}{c}{--} & \multicolumn{1}{c}{--} & \multicolumn{1}{c}{--} & \multicolumn{1}{c}{--} & \multicolumn{1}{c}{--} & \multicolumn{1}{c}{--} & \multicolumn{1}{c}{--} & \multicolumn{1}{c}{--} & \multicolumn{1}{c}{--} & \multicolumn{1}{c}{--} \\
\hline   
\end{tabular}
\normalsize

%\parbox{\hsize}{\textit{Notes}: The numbers use the same notation as Table~\ref{tab: coeffnu}.}
\end{table*}
%\end{sidewaystable}
\vspace{-3mm}
%\begin{sidewaystable}
\begin{table*}[ht]
\caption{Coefficients of equation (\ref{eq: poly}) and (\ref{eq: fnew}), for $\widetilde{f} = \widetilde{M}$ and $b_n$ calculated from  Eq.~(\ref{eq: solveb}).}    
\label{tab: coeffMb}
\centering                      
\tabcolsep=2.4pt
\footnotesize
\begin{tabular}{rrrrrrrrrrrr}
\hline \hline
$i$\textbackslash$j$& \multicolumn{1}{c}{0} & \multicolumn{1}{c}{1} & \multicolumn{1}{c}{2} & \multicolumn{1}{c}{3} & \multicolumn{1}{c}{4} & \multicolumn{1}{c}{5} & \multicolumn{1}{c}{6} & \multicolumn{1}{c}{7} & \multicolumn{1}{c}{8} & \multicolumn{1}{c}{9}  & \multicolumn{1}{c}{10} \\
\hline
0 & $ 7.075 \, (-4)$ & $ 1.030 \, (-4)$ & $-3.796 \, (-2)$ & $ 2.788 \, (-2)$ & $ 4.249 \, (-1)$ & $-6.912 \, (-1)$ & $-6.335 \, (-1)$ & $ 8.375 \, (-1)$ & $ 2.563 \ \ \ \ \ \ \ \, $ & $-4.113$ & $ 1.621$ \\
1 & $ 1.557 \, (-2)$ & $-4.107 \, (-2)$ & $-6.112 \, (-2)$ & $ 7.489 \, (-1)$ & $-1.700 \ \ \ \ \ \ \ \, $ & $-3.088 \, (-2)$ & $ 5.507 \ \ \ \ \ \ \ \, $ & $-8.413 \ \ \ \ \ \ \ \, $ & $ 5.055 \ \ \ \ \ \ \ \, $ & $-1.080$ & \multicolumn{1}{c}{--} \\
2 & $-3.518 \, (-2)$ & $ 5.740 \, (-2)$ & $ 2.328 \, (-1)$ & $-9.638 \, (-1)$ & $ 1.007 \ \ \ \ \ \ \ \, $ & $ 1.001 \ \ \ \ \ \ \ \, $ & $-3.228 \ \ \ \ \ \ \ \, $ & $ 2.706 \ \ \ \ \ \ \ \, $ & $-7.776 \, (-1)$ & \multicolumn{1}{c}{--} & \multicolumn{1}{c}{--} \\
3 & $ 1.986 \, (-2)$ & $-2.017 \, (-2)$ & $-1.276 \, (-1)$ & $ 3.000 \, (-1)$ & $-8.743 \, (-2)$ & $-4.228 \, (-1)$ & $ 5.203 \, (-1)$ & $-1.817 \, (-1)$ & \multicolumn{1}{c}{--} & \multicolumn{1}{c}{--} & \multicolumn{1}{c}{--} \\
4 & $ 1.274 \, (-3)$ & $-4.469 \, (-3)$ & $ 1.990 \, (-3)$ & $ 3.040 \, (-2)$ & $-7.734 \, (-2)$ & $ 7.556 \, (-2)$ & $-2.723 \, (-2)$ & \multicolumn{1}{c}{--} & \multicolumn{1}{c}{--} & \multicolumn{1}{c}{--} & \multicolumn{1}{c}{--} \\
5 & $-2.853 \, (-3)$ & $ 3.494 \, (-3)$ & $ 8.720 \, (-3)$ & $-1.959 \, (-2)$ & $ 1.656 \, (-2)$ & $-6.195 \, (-3)$ & \multicolumn{1}{c}{--} & \multicolumn{1}{c}{--} & \multicolumn{1}{c}{--} & \multicolumn{1}{c}{--} & \multicolumn{1}{c}{--} \\
6 & $-1.620 \, (-4)$ & $ 1.412 \, (-4)$ & $-4.977 \, (-5)$ & $-3.007 \, (-4)$ & $ 2.780 \, (-4)$ & \multicolumn{1}{c}{--} & \multicolumn{1}{c}{--} & \multicolumn{1}{c}{--} & \multicolumn{1}{c}{--} & \multicolumn{1}{c}{--} & \multicolumn{1}{c}{--} \\
7 & $ 2.557 \, (-4)$ & $-3.086 \, (-4)$ & $-2.993 \, (-4)$ & $ 3.039 \, (-4)$ & \multicolumn{1}{c}{--} & \multicolumn{1}{c}{--} & \multicolumn{1}{c}{--} & \multicolumn{1}{c}{--} & \multicolumn{1}{c}{--} & \multicolumn{1}{c}{--} & \multicolumn{1}{c}{--} \\
8 & $ 1.608 \, (-5)$ & $-2.209 \, (-6)$ & $-3.205 \, (-7)$ & \multicolumn{1}{c}{--} & \multicolumn{1}{c}{--} & \multicolumn{1}{c}{--} & \multicolumn{1}{c}{--} & \multicolumn{1}{c}{--} & \multicolumn{1}{c}{--} & \multicolumn{1}{c}{--} & \multicolumn{1}{c}{--} \\
9 & $-9.523 \, (-6)$ & $ 1.303 \, (-5)$ & \multicolumn{1}{c}{--} & \multicolumn{1}{c}{--} & \multicolumn{1}{c}{--} & \multicolumn{1}{c}{--} & \multicolumn{1}{c}{--} & \multicolumn{1}{c}{--} & \multicolumn{1}{c}{--} & \multicolumn{1}{c}{--} & \multicolumn{1}{c}{--} \\
10 & $-6.228 \, (-7)$ & \multicolumn{1}{c}{--} & \multicolumn{1}{c}{--} & \multicolumn{1}{c}{--} & \multicolumn{1}{c}{--} & \multicolumn{1}{c}{--} & \multicolumn{1}{c}{--} & \multicolumn{1}{c}{--} & \multicolumn{1}{c}{--} & \multicolumn{1}{c}{--} & \multicolumn{1}{c}{--} \\
\hline   
\end{tabular}
\normalsize

%\parbox{\hsize}{\textit{Notes}: The numbers use the same notation as Table~\ref{tab: coeffnu}.}
\end{table*}
%\end{sidewaystable}

\end{appendix}

\end{document}